\documentclass[10pt]{article}

\usepackage{amsmath}
\usepackage{amssymb}

\usepackage{graphicx}

\usepackage{cite}

\usepackage{color} 

\usepackage[labelfont=bf,labelsep=period,justification=raggedright]{caption}

\makeatletter
\renewcommand{\@biblabel}[1]{\quad#1.}
\makeatother

\date{}

\begin{document}

\begin{flushleft}
{\Large
\textbf{Aging in language dynamics}
}
\vspace{0.5cm}
\\
Animesh Mukherjee$^{1}$, 
Francesca Tria$^{1}$, 
Andrea Baronchelli$^{2}$,
Andrea Puglisi$^{3,4}$,
Vittorio Loreto$^{4,1\ast}$
\vspace{0.5cm}
\\
\bf{1} Institute for Scientific Interchange (ISI), Viale Settimio Severo 65, 10133 Torino, Italy
\\
\bf{2} Departament de Fisica i Enginyeria Nuclear, Universitat Politecnica de Catalunya, Campus Nord B4, 08034 Barcelona, Spain
\\
\bf{3} CNR-ISC Piazzale Aldo Moro 5, 00185 Roma, Italy
\\
\bf{4} Dipartimento di Fisica, ``Sapienza'' Universit\`a di Roma, Piazzale Aldo Moro 5, 00185 Roma, Italy
\\
$\ast$ E-mail: Vittorio.Loreto@roma1.infn.it
\end{flushleft}

\section*{Abstract}

Human languages evolve continuously, and a puzzling problem is how to
reconcile the apparent robustness of most of the deep linguistic
structures we use with the evidence that they undergo possibly slow,
yet ceaseless, changes. Is the state in which we observe languages
today closer to what would be a dynamical attractor with statistically
stationary properties or rather closer to a non-steady state slowly
evolving in time? Here we address this question in the framework of
the emergence of shared linguistic categories in a population of
individuals interacting through language games. The observed emerging
asymptotic categorization, which has been previously tested - with
success - against experimental data from human languages, corresponds
to a metastable state where global shifts are always possible but
progressively more unlikely and the response properties depend on the
age of the system. This aging mechanism exhibits striking quantitative
analogies to what is observed in the statistical mechanics of glassy
systems. We argue that this can be a general scenario in language
dynamics where shared linguistic conventions would not emerge as
attractors, but rather as metastable states.


\section*{Introduction}

A wide open question about the emergence and the evolution of shared
linguistic conventions concerns the role of
timescales~\cite{lieberman2007quantifying, pagel2007frequency}.
Phonetic, morphological, semantic, syntactic features of language vary
over time. A fair degree of variability of words and grammatical
structures can be observed in the diachronic study of a given
language~\cite{labov2007transmission}, and the proportions of
different linguistic variants used by individuals within a population
are not constant, but rather shift within
time~\cite{hruschka2009building}. An interesting case study is offered
by linguistic categories, the classical and prototypical example being
that of basic color
terms~\cite{berlin1991basic,lakoff-women,gardner1985msn,taylor2003lc},
or simply ``colors". For example, from Old to Middle English the
meaning of color terms shifted progressively from a brightness meaning
sense to the present-day hue sense~\cite{casson1997}, and similar
shifts have been documented for a wide array of
languages~\cite{maclaury1992brightness}. Also the variability existing
across different languages is an evidence of the continuous change of
linguistic categories, as pointed out by the data gathered in the
World Color Survey
(WCS)~\cite{berlin1991basic,cook2005world,berlinkay}. At the same
time, statistical analysis over a large number of languages have shown
that different color naming schemes share universal
patterns~\cite{kay2003rqc}, and, after a long debate, the existence of
universalities is nowadays widely
accepted~\cite{lakoff-women,gardner1985msn,taylor2003lc}.

General principles of categorization~\cite{garner1974processing} have
been claimed to be sufficient to account for the observed
universality~\cite{jameson1997s}. For example, it has been
suggested~\cite{jameson2005grue} that the simple principle according
to which categories are constructed as to maximize similarity within
categories and to minimize it across categories could be responsible
for cross-linguistic similarity, and a quantitative analysis based on
this intuition has confirmed its validity~\cite{Regieretal2007}.
Alternatively, it has been hypothesized that weak perceptual biases
could do the
job~\cite{deacon1998symbolic,tomasello2005constructing,christiansen2008language},
and recent numerical simulations have shown that this might well be
the case~\cite{pnas_wcs_2010}. In any case, both hypotheses have to
deal with the existence of a large variability, which is reflected in
the acknowledgment of the existence of non-optimal
categorizations~\cite{Regieretal2007} and in the consideration of the
weakness of the advocated perceptual biases~\cite{pnas_wcs_2010}. But
if optimality is the leading principle, how can languages get stuck in
suboptimal categorization schemes? Or what does it mean that
perceptual biases are ``weak''? i.e., why are they not able to drive
the evolution towards the very same end? And, more in general, which
is the mechanism that allow languages to appear static while they are
yet evolving?

Here we focus on the emergence of shared categorization patterns in
the framework of the so-called Category Game (CG)~\cite{cg_pnas}, a
language game~\cite{wittgenstein53english} through which a population
of individuals establishes a shared categorization that quantitatively
reproduces the average correlation among different human languages as
measured in the World Color Survey experiment~\cite{pnas_wcs_2010}. A
detailed analysis of the CG dynamics reveals that the physics of
glassy systems~\cite{mezard1987sgt,cavagna2009supercooled} can be the
proper framework to formalize the intuition that languages change at
the same time because of and notwithstanding the fact that they are
the outcome of a collective
behavior~\cite{halliday76,halliday89,yamaguchi96}.  Languages are
described as metastable states of global agreement, reconciling the
evidence that they do continuously
evolve~\cite{milroy_1992,labov_94,labov_01,mufwene2001ecology,croft2000explaining}
and they are at the same time stable enough to be intelligible across
a population.  

The physics of a so-called glass-forming liquid is such that when
rapidly undercooled under its melting temperature, it looses it
ability to flow on experimental time-scales and freezes in an
amorphous state with huge rheological times, while the most stable
state - the crystal - is never
reached~\cite{mezard1987sgt,cavagna2009supercooled}. This slowing down
process can be quantified through the so-called relaxation time which
turns out to be proportional to the viscosity of the fluid. Despite
languages and glassy systems stand apparently very far apart, it is
very intriguing to explore the analogy between a linguistic system and
a dynamical system featuring glassy properties.

Within this perspective, we study the dynamics of the CG model with
the tools of glass theory. In particular, we focus on the main aspects
which are peculiar to these physical systems, i.e., the scaling of
relaxation times and correlation functions with the population size as
well as with the age of the system. In particular, the larger is the
time over which one observes the system, formally known as the waiting
time, the slower will be its response, i.e., its ability to undergo
large-scale changes. From this perspective, the Category Game exhibits
a glassy behavior and constitutes a first quantitative evidence of a
very interesting link between cognitive science and the physics of
glassy systems.

\section*{Results}

\begin{figure}[!ht]
\begin{center}
 	\includegraphics[width=0.7\columnwidth,angle=0]{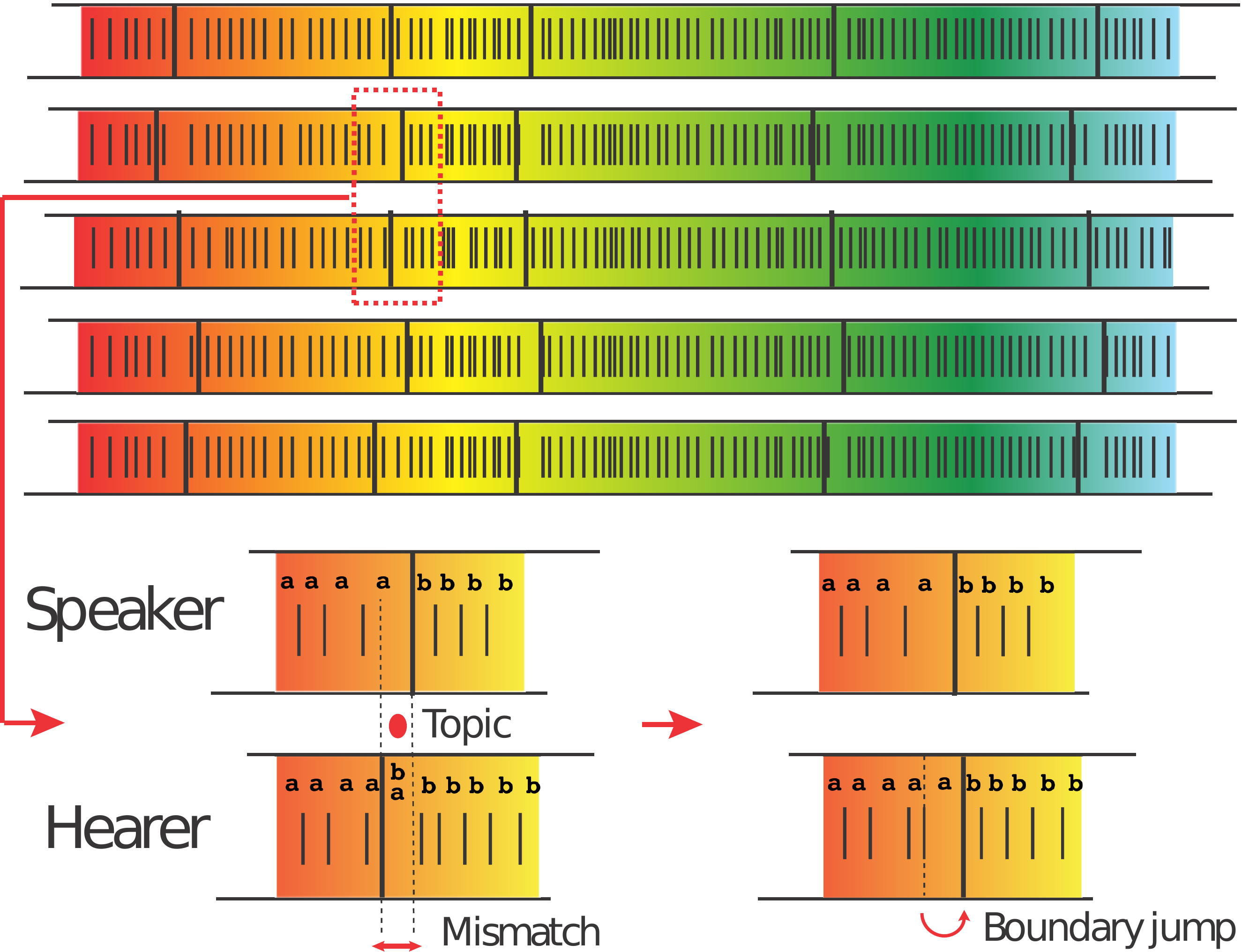}
\end{center} 	
  \caption{\label{fig1} {\bf Typical long-time configuration of five
      representative agents in the population.} For each agent
    perceptual and linguistic categories (separated by short and long
    bars, respectively) are shown. The highlighted portion of two
    agents illustrates an instance of a successful game in a so-called
    mismatch region between the linguistic categories of the two
    agents associated with the words ``a'' and ``b'' (see Materials
    and Methods for details). The hearer - in a previous game -
    learned the word ``a'' as a synonym for the perceptual category at
    the leftmost boundary of the linguistic category ``b''. During the
    game the speaker utters ``a'' for the topic; as a result the
    hearer deletes ``b'' from her inventory, keeping ``a'' as the name
    for that perceptual category, moving {\em de facto} the linguistic
    boundary.}
\end{figure}

The Category Game~\cite{cg_pnas} (see Materials and Methods for the
details) describes the emergence of a hierarchical category structure
made of two distinct levels: a basic layer, responsible for fine
discrimination of the environment (perceptual categories), and a
shared linguistic layer that groups together perceptions to guarantee
communicative success (linguistic categories). At each time step a
pair of individuals (one will be denoted as the speaker and the other
as the hearer) is randomly selected from the population to play a
language game that allow them to co-evolve the structure of their
categories as well as their form-meaning inventories. Fig.~\ref{fig1}
depicts a typical long-time configuration of the emerging category
structure. While the number of perceptual categories (separated by
short bars in fig. ~\ref{fig1}) is tuned by a parameter of the model
(see Materials and Methods) and can be arbitrarily large, the number
of linguistic categories (separated by long bars in fig. ~\ref{fig1}
and grouping together several perceptual categories sharing the same
word) turns out to be finite and small, as observed in natural
languages (for instance, like the basic color names across languages).

In the following, we report three sets of measures to establish the
emergence of aging in the Category Game.

\subsection*{Persistence of the Linguistic Categories}

\begin{figure}
\begin{center}
  \includegraphics[width=0.7\columnwidth,angle=0]{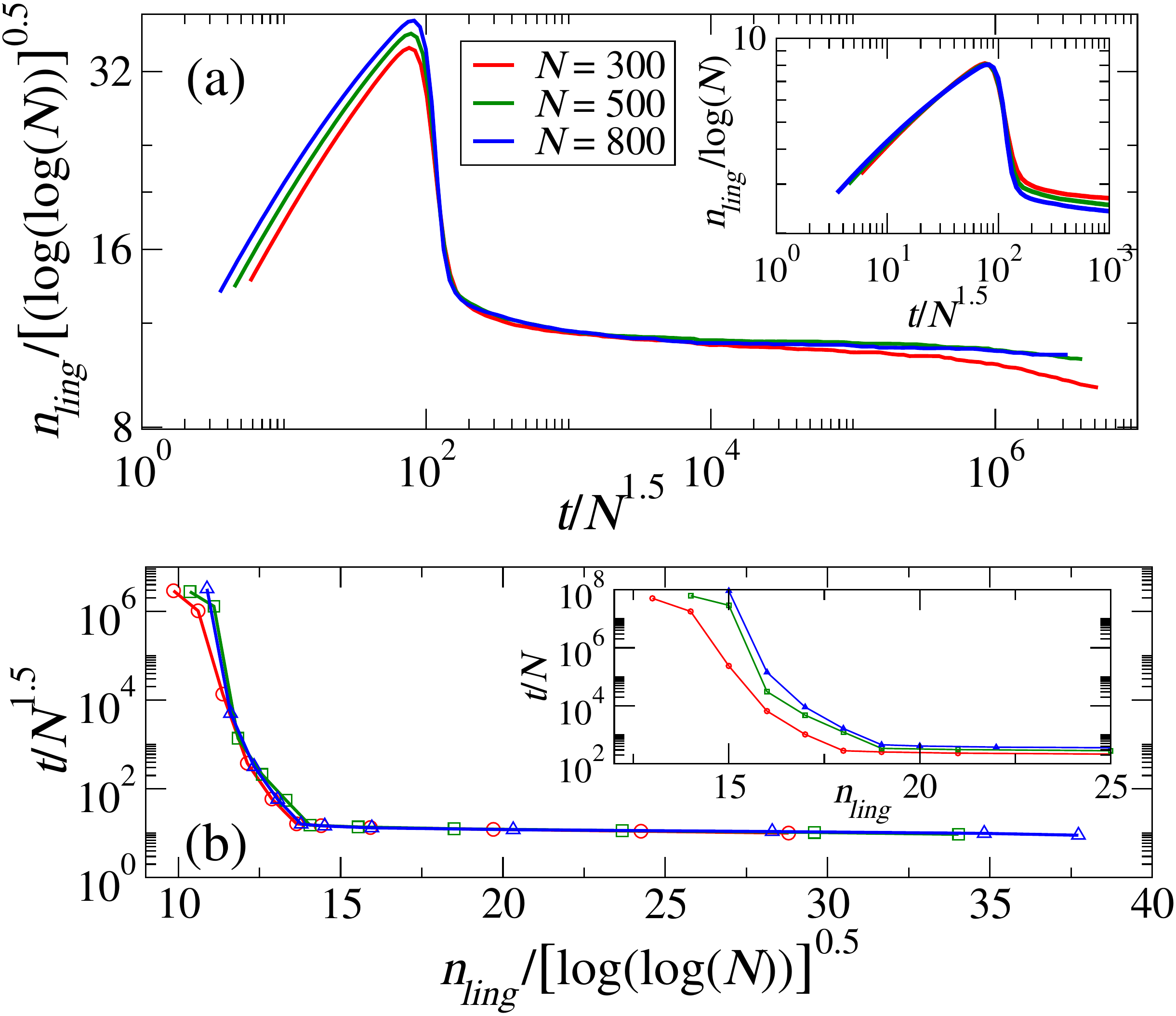}
 \end{center}
  \caption{\label{fig2} {\bf Persistence of the linguistic
      categories.} (a) The rescaled average number of linguistic
    categories $n_{ling}$ versus the rescaled number of games for
    three different population sizes ($N =300$, $500$ and $800$). The
    plateau behaviour for the average number of linguistic categories
    is collapsed by rescaling the ordinate by $\sqrt{\log{\log(N)}}$
    and the abscissa by $N^{3/2}$. The inset shows the data collapse
    for the first part of the evolution where the ordinate is rescaled
    by $\log{N}$ and the abscissa by $N^{3/2}$. (b) The rescaled
    persistence time of $n_{ling}$ (i.e., the time spent by the system
    in a configuration corresponding to an average of $n_{ling}$
    linguistic categories) versus the rescaled $n_{ling}$ for $N =
    300$, $500$ and $800$ (legends correspond to those in (a) except
    that the curves are plotted with both lines and symbols here).
    Once again the ordinate is rescaled by $N^{3/2}$ and the abscissa
    by $\sqrt{\log{\log(N)}}$ for data collapse. The inset shows a
    zoomed and uncollapsed version of the data (indicating the need
    for the collapse). Here the value of $d_{min}$ is set to the
    average human JND $0.0143$~\cite{long2006ssn}.}
 \end{figure}

We start by investigating the dynamics of the number of linguistic
categories emerging in the repertoire of each individual in a
population. Two regimes are clearly distinguished (fig.~\ref{fig2}a):
initially, corresponding to a series of uncorrelated games, the
average number of linguistic categories per individual $n_{ling}$
exhibits a rapid growth due to the pressure of discrimination (for a
detailed description of CG we refer to the Materials and Methods section), followed by a rapid drop due to the onset of consensus and
the merging of perceptual categories. A second regime is characterized
by a quasi-arrested dynamics signaled by a ``plateau'' region,
corresponding to a value of the average number of linguistic
categories of the order of ten~\cite{cg_pnas,pnas_wcs_2010}.
Interestingly, the dependence of the number of linguistic categories
on the population size $N$ is different in the two regimes. In the
first one, the average number of linguistic categories scales with
$\log{N}$ (see inset of fig.~\ref{fig2}a), while in the second regime the
dependence of the height of the plateau on the population size is
extremely weak (O($\sqrt{\log{\log{(N)}}}$)): the average number of
linguistic categories in the population remains limited to a small
value (of the order of $20 \pm 10$) even for very large population
sizes (up to billions of individuals). Furthermore, in the first
regime we recover a time dependence on the population size of order
$N^{3/2}$ (fig.~\ref{fig2}a), with a similar behaviour as in the Naming
Game~\cite{steels1995,baronchelli2008dan}, while the length of the
plateau features a much stronger dependence on $N$, reaching a
practically infinite value for large populations. At very large times, when the population is finite, the
average number of linguistic categories starts to drop. We shall come
back to this finite-size effect later on in the article. Most importantly, at the onset of the plateau region we
observe a slowing down of the dynamics signaled by the divergence of
the persistence time (fig.~\ref{fig2}b). The plateau region is thus the
interesting regime describing the persistence and evolution of the
category system, and we will next describe its properties by looking
at a more sophisticated dynamical quantity.

\subsection*{Autocorrelation function and metastability}

\begin{figure}
\begin{center}
  \includegraphics[width=0.65\columnwidth,angle=0]{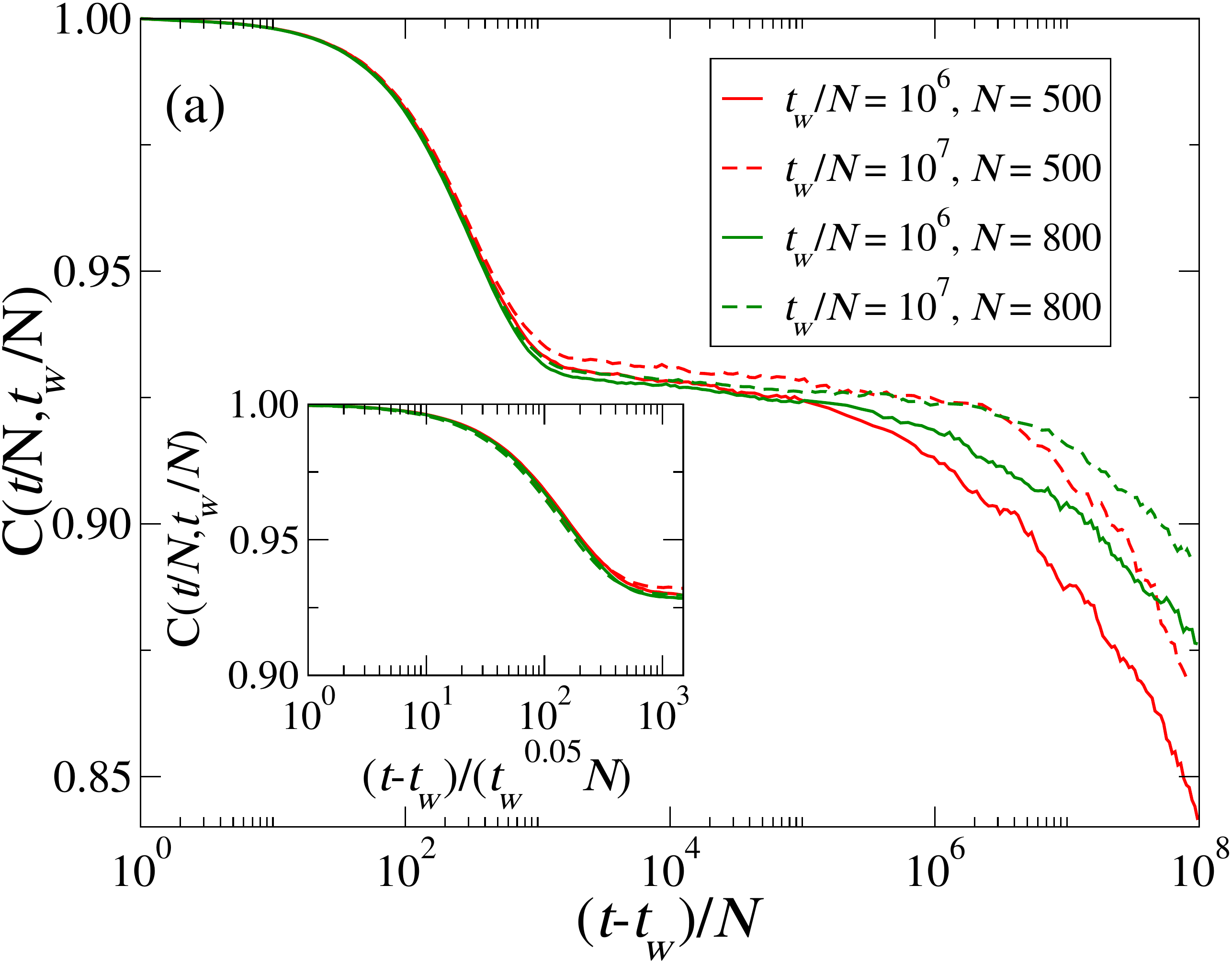}
  \includegraphics[width=0.65\columnwidth,angle=0]{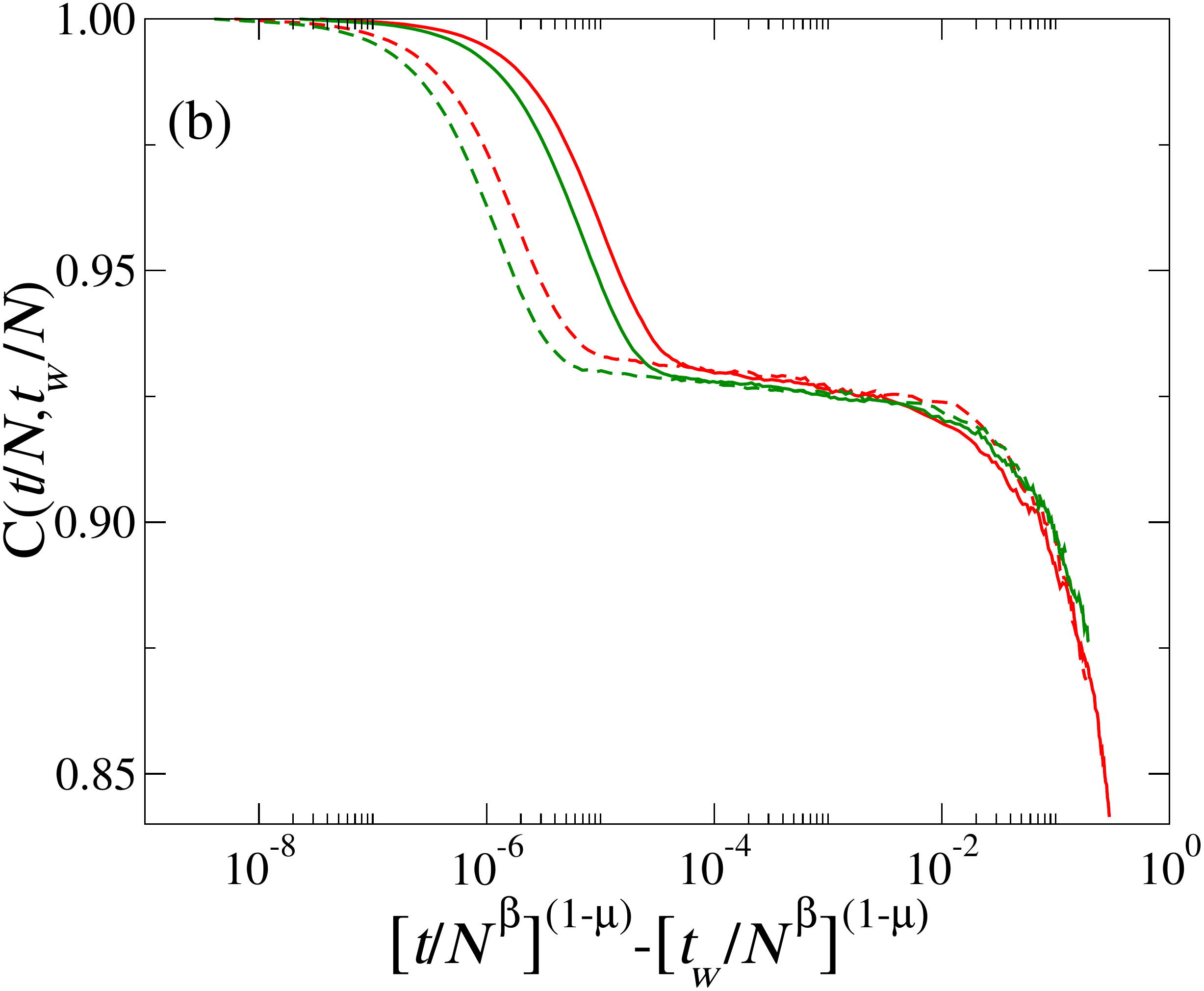}
\end{center}
  \caption{\label{fig3} {\bf Relaxation of the correlation functions.}
    (a) The autocorrelation $C(t/N, t_w/N)$ for $t_w/N = 10^6$, $10^7$
    and $ N = 500$, $800$. The inset shows the collapse of the $\beta$
    relaxation regime. In this regime, there is a very weak violation
    of the dependence of $C(t/N, t_w/N)$ on $t-t_w$ (time-translation
    invariance). (b) The collapse of the autocorrelation functions
    shown in (a) in the $\alpha$ relaxation regime indicating
    sub-aging ($\mu \simeq 0.75$). This result shows that the
    relaxation is strongly dependent on the size of the population
    ($\sim N^{\beta}$ with $\beta\simeq 4.5$). Here again the value of
    $d_{min}$ is set to the average human JND
    $0.0143$~\cite{long2006ssn}.}
 \end{figure}

A system is said to be in dynamical equilibrium when it shows
invariance under time translations; if this holds, any observable
comparing the system at time $t_w$ with the system at time $t_w+t$
does not depend on $t_w$. In contrast, a system undergoing aging is
not invariant under time translation, i.e., time is not homogeneous.
This property can be revealed by measuring correlations of the system
at different times. Here we consider a suitably defined
autocorrelation function, which we term $C(t, t_w)$: at
time $t_w$ we save a copy of the configuration of all the agents in
the population and subsequently, at time instances greater than $t_w$,
we compute the linguistic overlap of each agent with its copy saved at
$t_w$; finally, we average this quantity over all agents (see Materials and Methods for detailed
definitions). Results are presented in fig.~\ref{fig3} for two different
population sizes. We recognize two different time scales, which we can
associate to local or individual (fast) and collective or
population-related (slow) dynamics. In particular, for $(t-t_w)/N<
10^3$, $C(t, t_w)$ depends (almost) only on $t-t_w$ (see inset of
fig.~\ref{fig3}a). This phenomenon corresponds to what is known in the physics
of glassy systems as the $\beta$-relaxation regime. This fast dynamics
corresponds to the microscopic dynamics of the boundaries between
linguistic categories at the individual level.

On the other hand, for $(t-t_w)/N> 10^3$, $C(t, t_w)$ reveals aging,
corresponding to the so-called $\alpha$-relaxation regime in glassy
systems. This slow dynamics corresponds to the collective dynamics of
the boundaries between linguistic categories at the population level.
We find, in particular, for a fixed population size, a dependence on
$t_w$ of the form:

\begin{equation}
  C(t,t_w)=\tilde{C}(t^{1-\mu}-t_w^{1-\mu}) \, ,
\end{equation}

\noindent with $\mu=0.75$ (see fig.~\ref{fig3}b). Note that the same type of
dependence on $t$ and $t_w$ is also found in correlation functions of
real glasses~\cite{vincent1997,marco2001}, making quantitative the
analogy of the CG dynamics with the dynamics of real
out-of-equilibrium physical systems that exhibit sub-aging behavior.
Let us simply recall that in the limit $\mu \rightarrow 1$ one
recovers a pure aging behaviour since $t^{1-\mu}-t_w^{1-\mu}
\rightarrow log(t/t_w)$. In the opposite limit, $\mu \rightarrow 0$, one
recovers invariance under time translations.

Fig.~\ref{fig3} also reveals that the dependence of the length of the plateau on the population size is of the order of $N^{\beta}$ with $\beta\simeq 4.5$. This suggests that the attractor of the dynamics, where a single linguistic
category spreads over the whole interval, is practically never
reached for large enough population sizes, and metastable states with
a limited number of linguistic categories last for a practically
infinite time.

\subsection*{Finite-size effects}

\begin{figure}
\begin{center} 
  		 \includegraphics[width=0.7\columnwidth,angle=0]{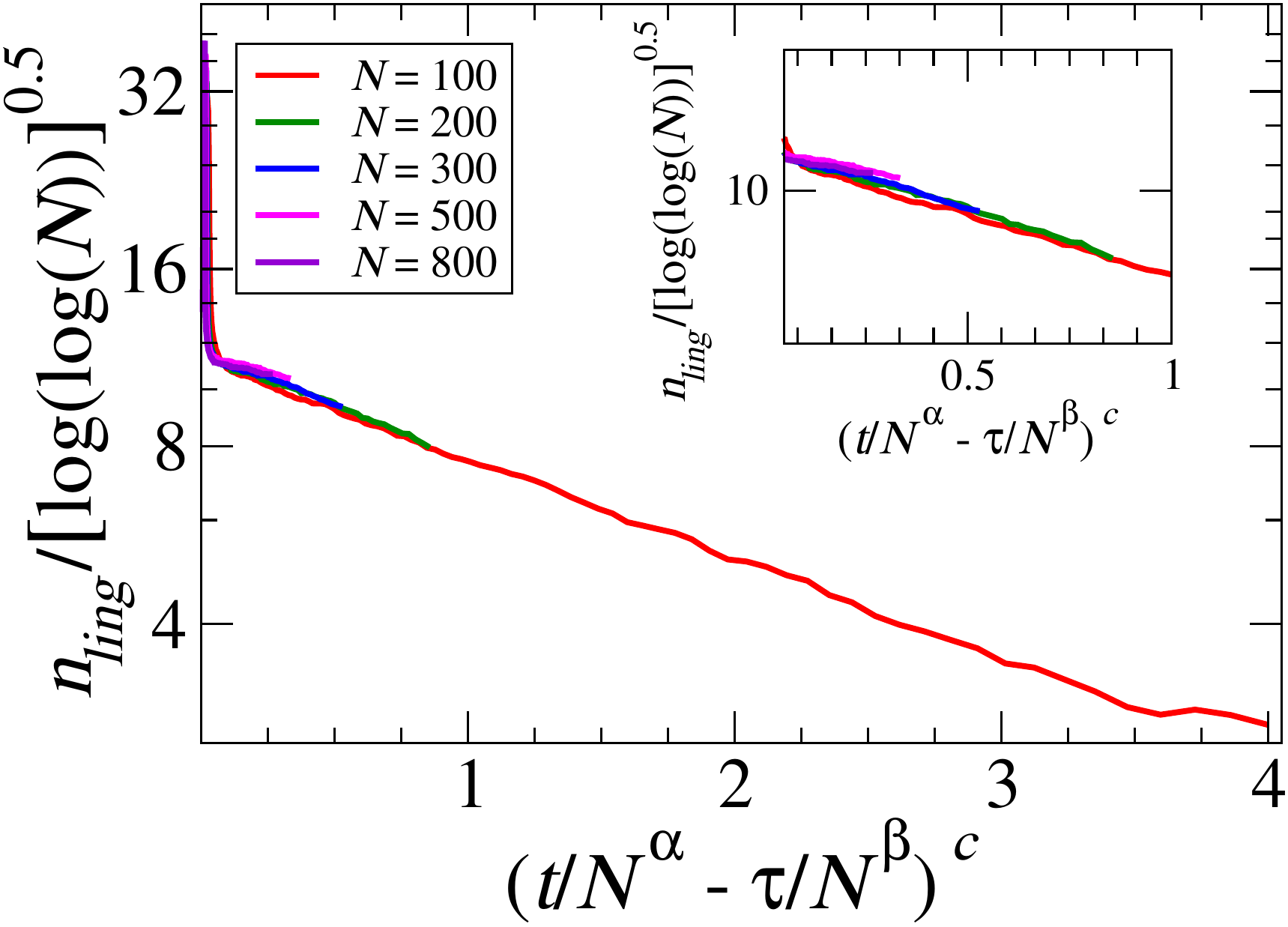}
   \end{center}
  \caption{\label{fig4} {\bf Finite-size effects.} The rescaled
    average number of linguistic categories $n_{ling}$ versus the
    rescaled number of games for five different population sizes ($N
    =100$, $200$, $300$, $500$ and $800$). The bending region of the
    curves is collapsed by rescaling the number of linguistic
    categories by $\sqrt{\log{\log(N)}}$ and the time axis as
    $(t/N^{\alpha} - \tau/N^{\beta})^{c}$ where $\alpha \simeq 3$,
    $\beta \simeq 4.5$, $\tau \simeq 2.87\times 10^5$ and $c \simeq
    1/3$. The inset shows a zoomed version of the same plot to present
    a better visualization of the data collapse.}
 \end{figure}

In this section, we consider finite-size effects. We focus, in
particular, on the behaviour of the average number of linguistic
categories as a function of time (as observed in fig.~\ref{fig2}a). A careful observation reveals that for very long times the
plateau behaviour leaves room for a bending of the curves leading to a
reduction in the average number of linguistic categories. This bending
occurs earlier for small populations, i.e., for small system sizes.
Fig.~\ref{fig4} shows the collapse of the curves for the average number of
linguistic categories for different system sizes from $N=100$, for
which the bending is stronger, to $N=800$. The collapse is aimed at
superimposing only the bending region. It turns out that one collapses
the curves after a rescaling of the abscissa as $t \rightarrow
b(t/N^{\alpha} - \tau/N^{\beta})^{c}$, where the term $\tau/N^{\beta}$
allows to superimpose the onset of the bending region while the term
$t/N^{\alpha}$ is the time rescaling well inside the bending region
and $b$ is a constant. The value of $\beta\simeq 4.5$ is consistent
with what is observed in the collapse of the correlation function (see
fig.~\ref{fig3}) and confirms the idea that the length of the
plateau region is scaling with a large power of the system size. On
the other hand the bending region exhibits a characteristic time
scaling as $N^{\alpha}$ with $\alpha\simeq 3$ and the overall
behaviour is well fitted by a stretched exponential function with an
exponent $c$ close to $1/3$.

\begin{figure}
\begin{center}
 	\includegraphics[width=0.7\columnwidth,angle=0]{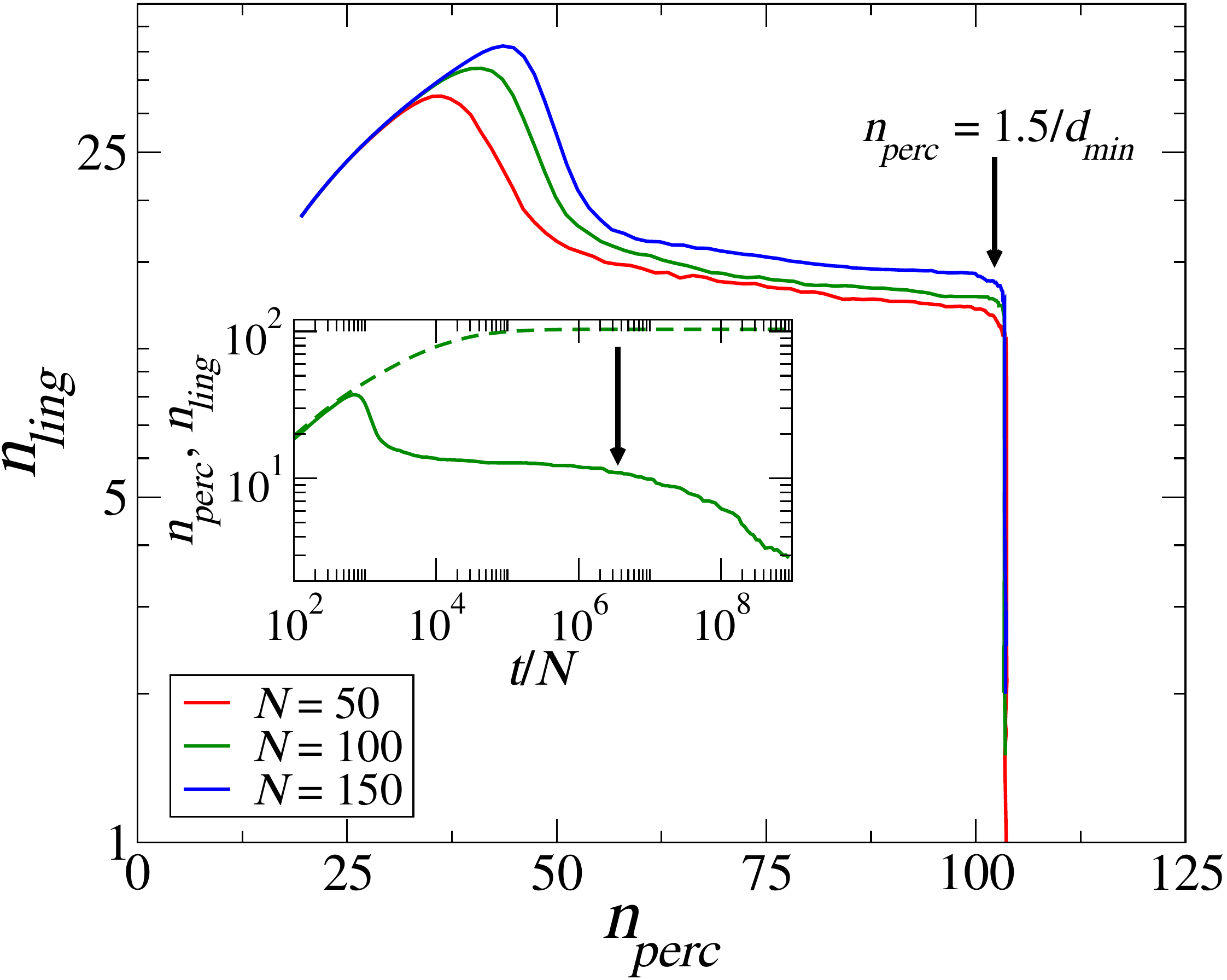}
\end{center}
  \caption{\label{fig5} {\bf Linguistic vs. perceptual categories}.
    Parametric plot of the number of linguistic categories vs. the
    number of perceptual categories, $n_{perc}$, for different
    population sizes for which the bending region is accessible within
    a reasonable time ($N = 50$, $100$ and $150$). It is evident that
    there is a transition (indicated by the bold arrow) between a
    long-lasting regime where the number of perceptual categories
    keeps increasing, though at a very slow pace, and a regime where
    discrimination stops, the number of perceptual categories does not
    increase anymore and one observes only a decrease in the number of
    linguistic categories. The inset shows one representative example
    of the time evolution of $n_{perc}$ and $n_{ling}$ for $N$ = 100
    where the bold arrow marks the onset of the bending.}
\end{figure}

 The onset of the bending region is marked by a clear phenomenon
 occurring in the structure of the perceptual and linguistic
 categories. In the plateau region discrimination keeps taking place,
 though at a slow pace, while at the onset of the bending region
 discrimination ceases and one is left with a pure dynamics of the
 boundaries between linguistic categories (see fig.~\ref{fig5}). A detailed
 description of the dynamics of the domain boundaries in the plateau
 and in the bending region is out of the scope of the present paper
 and it will be presented elsewhere. It is nevertheless interesting to
 mention that one can describe the dynamics of the domain boundaries
 between linguistic categories in terms of correlated random walkers.
 The crucial difference between the plateau region (where aging
 occurs) and the bending region is that when the system ages the
 number of perceptual categories, which represent the underlying
 lattice where the random walkers can diffuse, is an increasing
 function of time while it is a constant when finite-size effects
 start.

\section*{Discussion}

In summary, our {\em in silico} experiment demonstrates a strong analogy between
the slow dynamics of the Category Game and that of a suddenly quenched
glass-former, in many different crucial aspects. The relaxation time
of the number of linguistic categories shows a huge increase, similar
to a singularity, at a finite number of categories: the system traps
itself in a metastable state and the dynamics appears arrested even
though the final state (only one category) is far from being reached.
The dependence of the number of linguistic categories on the
population size appears to be extremely weak (slower than a
logarithm), and this could account for the universality of the number
of color names (between $2$ and $11$) among different languages. The
decorrelation of the system, apart from being slow, is also
age-dependent, thus suggesting a possible explanation for the
existence of more stable conserved properties of a language.

Taking a wider perspective, our results suggest that glassy behavior
could provide important conceptual and technical tools to address the
general problem of language change from a new perspective.
Metastability, for example, allows to unfold the seemingly paradoxical
nature of language change, according to which languages evolve because
they are spoken by a large number of speakers but the evolution is
frustrated since the speakers are indeed many. These intuitions are of
course not a novelty (see for
instance~\cite{mufwene2001ecology,croft2000explaining}), but for the
first time they have been properly quantified, in a numerical model
closely connected to experimental data, and addressed in a well
established framework. Future work could take into account crucial
phenomena like language contact or more general cultural evolution
processes, thus paving the way for a comparison with true historical
data.

\section*{Materials and Methods}

\subsubsection*{The Category Game}

The computational model used for this study, introduced in
\cite{cg_pnas}, investigates how a population of individuals can
develop a shared repertoire of linguistic categories, i.e., co-evolve
their own system of symbols and meanings, without any pre-defined
categorization and by only means of elementary language
games~\cite{wittgenstein53english}. A population of $N$ artificial
agents is considered and to each agent (or individual) a continuous
perceptual space (e.g., the visible light spectrum) is associated
which, without any loss of generality, is assumed to be the interval
$[0, 1)$. A categorization pattern refers to a partition of this
interval into sub-intervals, or perceptual categories. Each individual
has a dynamical form-meaning repertoire linking perceptual categories
(meanings) to words (forms) representing their linguistic counterpart.
The perceptual categories and the words associated with them co-evolve
through a sequence of elementary language games among the agents.
Initially, all individuals have a single perceptual category $[0, 1)$
and no name associated to it.

At each time step a pair of individuals (one will be denoted as the
speaker and the other as the hearer) is randomly selected from the
population and based on the success or failure of communication, both
rearrange their form-meaning inventories. Both the speaker and the
hearer are presented with a scene made of $M \ge 2$~\footnote{Without
  any loss of generality we will use in all the simulations $M=2$.}
stimuli (objects), where a stimulus is a real number in the interval
$[0, 1)$. Any two objects in the scene cannot appear at a distance
closer than $d_{min}$: this is the only parameter of the model, fixing
a minimal length scale which encodes a non infinite resolving power of
any perception, for instance, the human Just Noticeable Difference
(see below) in the case of colors. One of the objects is randomly
selected to be the {\em topic} of the game and is known only to the
speaker. The speaker checks whether the topic is the unique stimulus
corresponding to one of its perceptual categories. If the two stimuli
fall in one perceptual category, then the category is divided into two
new categories by a barrier located in the center of the segment
connecting the two stimuli. Both the new categories inherit the words
associated to the original category plus a new word; this process is
termed as {\em discrimination}. Subsequently, the speaker utters the
most relevant name of the category containing the topic, where the
most relevant name corresponds to either the last name used in a
winning game or the new name in case the category has just been
created. If the hearer does not have a category with this name, the
game is a failure. If the hearer recognizes the name and has any
object in one or more categories associated with that name, then it
picks randomly one of these objects. If the object picked is the
topic, then the game is a success; otherwise, it is a failure. In case
of failure, the hearer learns the name used by the speaker for the
category corresponding to the topic. In case of success, that name
becomes the most relevant for that category and all other competing
names are removed from the inventory associated with the category for
both the players. An example of the evolving dynamics is shown in fig.~\ref{fig1}.

It is worth to remark that in the framework of the CG model it was
recently possible to reproduce the outcomes of the World Color Survey
(WCS)~\cite{pnas_wcs_2010}. This is a first evidence that the model
can account for the {\em universality} of categorization patterns
across cultures by means of only weak constraints on the perceptual
space of the individuals. In~\cite{pnas_wcs_2010}, universal
categorization patterns have been identified among populations whose
individuals are endowed with the human Just Noticeable Difference
(JND) function, describing the resolution power of the human eye to
variations in the wavelength of the incident light~\cite{long2006ssn}.
In the simulations presented here, for the sake of simplicity, we
adopt a constant $d_{min}$ equal to the average human JND ($0.0143$),
after having checked that the dynamical properties do not depend on
eventual modulations of the JND function.

\subsubsection*{A fast implementation of the Category Game}

In this paper, for the investigation of the long-time dynamics of the model we
devised and adopted a fast version of the Category Game that we
briefly describe here, referring to a forthcoming paper for further
details. The central idea behind the fast algorithm we implement is
that of avoiding all the unnecessary games without outcomes (i.e.,
without changes in the configuration of either the speaker or the
hearer or both) by forcing that each two players' game has an outcome
and rescaling time accordingly. In this way, we do not alter the
original dynamics since we conserve the playing order of the pairs
speaker-hearer as well as the probability of playing in a given
region.

More concretely, at each step we extract one pair speaker-hearer
according to the probability $P_{out}$ that an outcome will result if
the pair plays, where an outcome is defined as any change in the
repertoire of the speaker and/or the hearer\footnote{In the considered
  dynamics, we found that the probabilities $P_{out}$ of all the
  pairs, at each given time, followed a peaked distribution, so that
  we could ultimately randomly extract one pair at each step without
  altering the results and significantly reducing the computational
  complexity.}. Furthermore, the region for placing the topic is
extracted according to the probability that this choice will produce
an outcome, and the object is extracted consistently. After each game
time is increased by $1/P_{out}$. In this way time becomes a dependent
variable. We checked that this fast implementation of the Category
Game features the same dynamical properties of the original model for
all the quantities of interest.

\subsubsection*{Dynamical properties of the Category Game}

\begin{figure}
\begin{center}
  \includegraphics[width=0.7\columnwidth,angle=0]{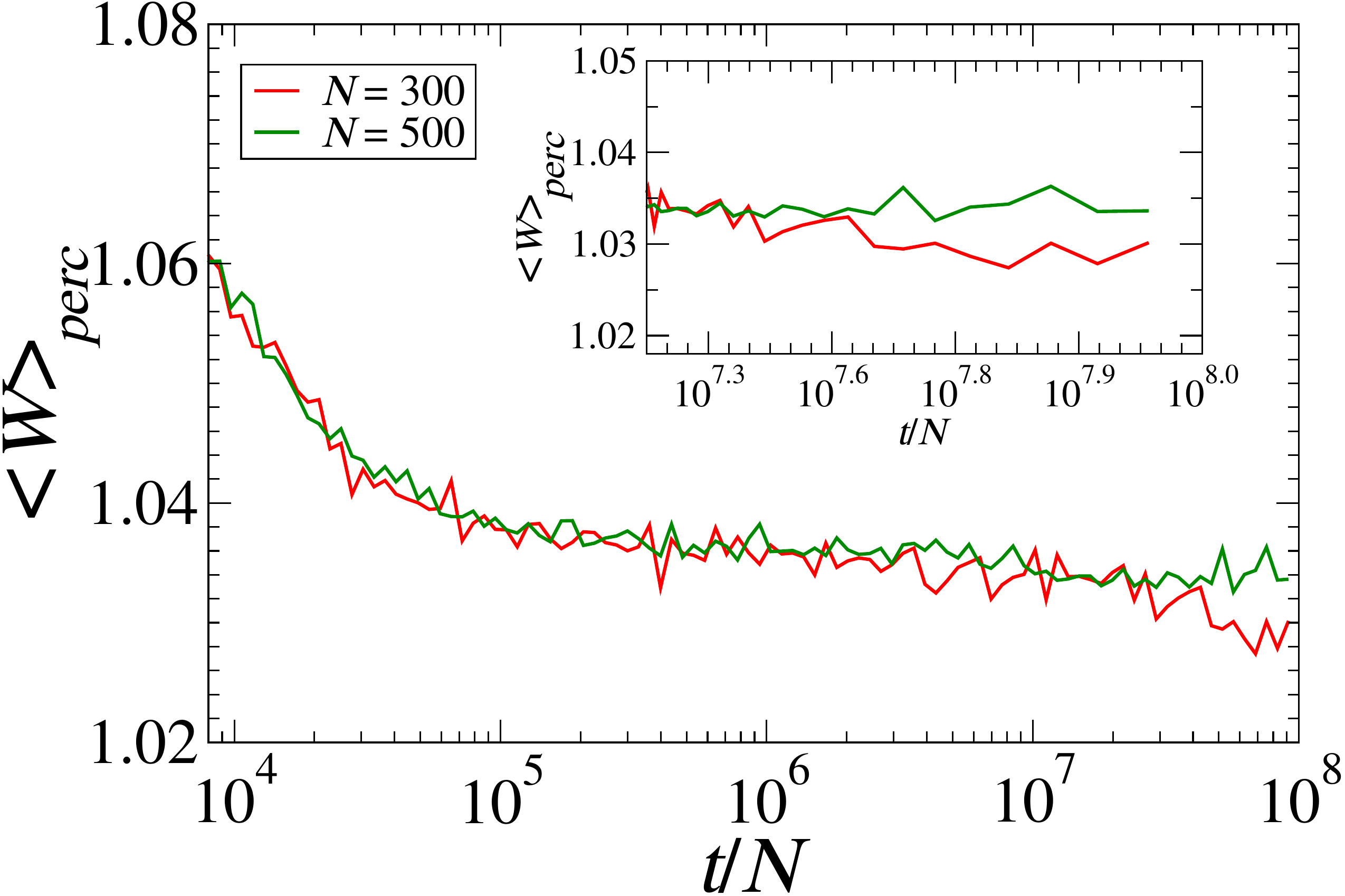}
\end{center}
  \caption{\label{fig6} {\bf Words per perceptual category.} The
    average number of words per perceptual category $\langle W
    \rangle_{perc}$ across the population of $N$ = 300, 500 agents
    versus the number of games per player. The inset is a zoom showing
    $\langle W \rangle_{perc}$ after $10^7$ games per player. Clearly,
    $\langle W \rangle_{perc}$ does not settle to one even after a
    very long time. The value of $d_{min}$ here is equal to $0.0143$
    which is the average of human JND (when projected on the $[0, 1)$
    interval)~\cite{long2006ssn}.}
 \end{figure}

In the Category Game dynamics it is possible to distinguish two
different phases. In the first regime, the number of perceptual
categories increases due to the pressure of discrimination, and at the
same time many different words are used by different agents for naming
similar perceptual categories. This kind of synonymy reaches a peak
and then drops~\cite{cg_pnas} in a fashion similar to the well-known
Naming Game~\cite{steels1995,baronchelli_ng_first,baronchelli2008dan}.
A second phase starts when most of the perceptual categories are
associated with only one word (see fig.~\ref{fig6}). During this phase, words
are found to expand their dominion across adjacent perceptual
categories. In this way, sets of contiguous perceptual categories
sharing the same words are formed, giving raise to what we define as
{\em linguistic categories} (see fig.~\ref{fig1}). An
important outcome thus is the emergence of a hierarchical category
structure made of two distinct levels: a basic layer, responsible for
fine discrimination of the environment, and a shared linguistic layer
that groups together perceptions to guarantee communicative success.
Remarkably, the emergent number of linguistic categories in this phase
turns out to be finite and small~\cite{cg_pnas}, as observed in
natural languages, even in the limit of an infinitesimally small
length scale $d_{min}$, as opposed to the number of the underlying
perceptual categories which is of order $1/d_{min}$.

\subsubsection*{Linguistic overlap and autocorrelation function}

The autocorrelation function $C(t, t_w)$ is defined as the average in the population of the individual
linguistic autocorrelation, which, in turn, is defined as
 the overlap of the linguistic categories~\cite{cg_pnas} of the
 considered individual at time $t_w$, with itself at a later time
 $t$. It then reads:
\begin{equation}\label{eq:ov}
  C(t,t_w) = \sum_{i} \frac{O_{i}(t,t_w)}{N} \textrm{ with } 
  O_{i}(t,t_w) = \frac{2\sum_{c_{i(t)}^{i(t_w)}} {(l_{c_{i(t)}^{i(t_w)}})}^2}
  {\sum_{c_{i(t)}} {(l_{c_{i(t)}})}^2 + \sum_{c_{i(t_w)}} {(l_{c_{i(t_w)}})}^2}
\end{equation}
\noindent 
where $l_c$ is the width of the linguistic category $c$, $c_{i(t)}$ is a
linguistic category of the $i^{th}$ agent at time $t$ and $c_{i(t)}^{i(t_w)}$ is the generic
category of the intersection set containing all of the linguistic category boundaries of
the agent $i$ at time $t$ and its previous image  saved at $t_w$. The function
$O_{i} (t,t_w)$ returns a value proportional to the degree of alignment of
the two category inventories reaching its maximum unitary value when
they are perfectly aligned.

\section*{Acknowledgments}

A. Baronchelli acknowledges support from the Spanish Ministerio de
Ciencia e Innovaci\'{o}n through the Juan de la Cierva program, as
well as from project FIS2007-66485-C02-01 (Fondo Europeo de Desarrollo
Regional), and from the Junta de Andaluc\'{i}a project P09-FQM4682. A.
Puglisi acknowledges support by the Italian MIUR under the FIRB-IDEAS
grant RBID08Z9JE.

\bibliographystyle{plos2009}
\bibliography{plos_bib}

\newpage

\section*{Figures and Legends}

\end{document}